\documentclass[sigconf,authorversion]{acmart}

\usepackage{subfig}
\usepackage{tabularx}
\usepackage{booktabs}
\usepackage{multirow}
\usepackage[acronym]{glossaries}

\copyrightyear{2021}
\acmYear{2021}
\setcopyright{acmlicensed}\acmConference[ARES 2021]{The 16th International Conference on Availability, Reliability and Security}{August 17--20, 2021}{Vienna, Austria}
\acmBooktitle{The 16th International Conference on Availability, Reliability and Security (ARES 2021), August 17--20, 2021, Vienna, Austria}
\acmPrice{15.00}
\acmDOI{10.1145/3465481.3465743}
\acmISBN{978-1-4503-9051-4/21/08}

\newcommand{\groupplus}[0]{\textsc{+atk}}
\newcommand{\letter}[0]{\textsc{letter}}
\newcommand{\mail}[0]{\textsc{email}}  
\newcommand{\letterplus}[0]{\textsc{letter}\groupplus{}}
\newcommand{\mailplus}[0]{\textsc{email}\groupplus{}}  
\newcommand{\control}[0]{\textsc{control}}

\newcommand{\phpinfo}{PHP\-Info}

\newcommand{\mailmedian}[0]{39.3}
\newcommand{\mailplusmedian}[0]{40.4}
\newcommand{\lettermedian}[0]{64.3}
\newcommand{\letterplusmedian}[0]{51.9}
\newcommand{\allmailmedian}[0]{40.0}
\newcommand{\alllettermedian}[0]{59.0}
\newcommand{\allplainmedian}[0]{52.0}
\newcommand{\allplusmedian}[0]{45.3}
\newcommand{\allnotifiedmedian}[0]{48.9}
\newcommand{\controlmedian}[0]{4.3}
\newcommand{\statusmedian}[0]{45.9}
\newcommand{\vcsmedian}[0]{40.8}
\newcommand{\databasemedian}[0]{60.0}
\newcommand{\phpinfomedian}[0]{48.3}

\begin{document}

\title[Snail Mail Beats Email Any Day: On Effective Operator Security Notifications in the Internet]{Snail Mail Beats Email Any Day:\\On Effective Operator Security Notifications in the Internet}

\author{Max Maass}
\orcid{0000-0001-9346-8486}
\affiliation{%
	\institution{Secure Mobile Networking Lab}
	\institution{Technical University of Darmstadt}
	\streetaddress{Pankratiusstraße 2}
    \city{Darmstadt}
	\postcode{64289}
	\country{Germany}
}
\email{mmaass@seemoo.tu-darmstadt.de}

\author{Marc-Pascal Clement}
\affiliation{%
	\institution{Secure Mobile Networking Lab}
	\institution{Technical University of Darmstadt}
	\streetaddress{Pankratiusstraße 2}
	\city{Darmstadt}
	\postcode{64289}
    \country{Germany}
}
\email{mclement@seemoo.tu-darmstadt.de}

\author{Matthias Hollick}
\orcid{0000-0002-9163-5989}
\affiliation{%
	\institution{Secure Mobile Networking Lab}
	\institution{Technical University of Darmstadt}
	\streetaddress{Pankratiusstraße 2}
	\city{Darmstadt}
	\postcode{64289}
	\country{Germany}
}
\email{mhollick@seemoo.tu-darmstadt.de}


\begin{abstract}
In the era of large-scale internet scanning, misconfigured websites are a frequent cause of data leaks and security incidents.
Previous research has investigated sending automated email notifications to operators of insecure or compromised websites, but has often met with limited success due to challenges in address data quality, spam filtering, and operator distrust and disinterest.
While several studies have investigated the design and phrasing of notification emails in a bid to increase their effectiveness, the use of other contact channels has remained almost completely unexplored due to the required effort and cost.
In this paper, we investigate two methods to increase notification success: the use of letters as an alternative delivery medium, and the description of attack scenarios to incentivize remediation.
We evaluate these factors as part of a notification campaign utilizing manually-collected address information from 1359 German website operators and focusing on unintentional information leaks from web servers.
We find that manually collected addresses lead to large increases in delivery rates compared to previous work, and letters were markedly more effective than emails, increasing remediation rates by up to 25 percentage points. 
Counterintuitively, providing detailed descriptions of possible attacks can actually \emph{decrease} remediation rates, highlighting the need for more research into how notifications are perceived by recipients.
\end{abstract}

\ccsdesc[500]{Security and privacy~Web application security}
\ccsdesc[100]{Security and privacy~Usability in security and privacy}

\keywords{web security, notification study, information leakage}

\maketitle

\glsdisablehyper
\newacronym{DKIM}{DKIM}{DomainKeys Identified Mail}
\newacronym{SPF}{SPF}{Sender Policy Framework}
\newacronym{TLD}{TLD}{Top-Level Domain}
\newacronym{GDPR}{GDPR}{General Data Protection Regulation}
\newacronym{CERT}{CERT}{Computer Emergency Response Team}
\newacronym{ISP}{ISP}{Internet Service Provider}
\newacronym{GA}{GA}{Google Analytics}
\newacronym{VCS}{VCS}{Version Control System}
\newacronym{TLS}{TLS}{Transport Layer Security}
\newacronym{HTTPS}{HTTPS}{Hypertext Transfer Protocol Secure}
\newacronym{SSH}{SSH}{Secure Shell}
\newacronym{OS}{OS}{Operating System}
\newacronym{DNS}{DNS}{Domain Name System}
\newacronym{DDoS}{DDoS}{Distributed Denial of Service}

\section{Introduction}\label{sec:intro}

Operating a website securely requires constant attention to keep both software and configurations up to date and to avoid security issues. It is inevitable that some operators will make mistakes, which can be dangerous and expensive. For example, the Equifax breach was caused by a missing software update \cite{ArsEquifax}, and even a well-publicized vulnerability like Heartbleed still saw 3\,\% of the Alexa Top Million websites remain vulnerable two months after disclosure~\cite{Durumeric2014}.

Researchers have attempted to send notifications to operators of insecure \cite{Durumeric2014,Li2016WWW,Stock2016,Stock2018}, compromised \cite{Vasek2012,Canali2013,Cetin2016,Cetin2018,Cetin2019}, or misconfigured \cite{Kuhrer2014,Li2016Usenix,Cetin2017,Zeng2019,Maass2021} systems, finding increased remediation rates, but also problems with undeliverable messages \cite{Stock2016,Stock2018,Cetin2016,Durumeric2014,Li2016Usenix,Cetin2017} and operator distrust \cite{Stock2018,Cetin2018,Cetin2019,Zeng2019}. They also observed large numbers of systems that remained unfixed, even after multiple notifications.

We investigate the role of two factors in a notification campaign: the \emph{medium} of the message, where we compare emails and postal letters, and the presence of \emph{attack scenarios} in the message, where we describe attacks enabled by the reported issues. We also seek to avoid the reachability issues reported by previous studies by manually collecting contact information for all websites, thereby operating with the highest quality of contact information available.
Our notifications also contain a link to a self-service tool where recipients receive additional information and can verify if their website is still vulnerable. 

We conduct our study with $N=1359$ operators of German websites suffering from unintentional information leakage that would allow attackers to gain access to detailed information about the software running on the server, cryptographic keys, or entire databases.
These issues are easily remediated, and their remediation will not lead to incompatibilities. This is crucial, as operators may be hesitant to migrate away from old, insecure software versions because this would break compatibility with other software \cite{vaniea2016}.

Our paper makes the following contributions:
\begin{itemize}
    \item We compare the effectiveness of letters and emails in a randomized controlled notification experiment, using manually-collected address information to operate under best-case data quality assumptions.
    \item We investigate the effect of adding or withholding details about how the vulnerabilities could be used in a realistic attack, to act as an incentive to remediate.
    \item We provide notification recipients with a self-service tool to evaluate if their remediation attempts were successful, and monitor its use.
\end{itemize}

The rest of this paper is structured as follows: after discussing previous studies in \autoref{sec:relatedwork}, we describe our experimental setup in \autoref{sec:methodology}. We give an overview about the obtained results in \autoref{sec:results} and offer an interpretation in \autoref{sec:discussion}. Finally, we conclude in \autoref{sec:conclusion}.

\section{Related Work}
\label{sec:relatedwork}
Previous work has conducted notification campaigns about a variety of issues, including the security of websites \cite{Vasek2012,Canali2013,Durumeric2014,Cetin2016,Li2016WWW,Stock2016,Stock2018,Zeng2019}, or \gls{DNS} servers \cite{Cetin2017}, misconfigured systems leading to \gls{DDoS} amplification \cite{Kuhrer2014,Li2016Usenix,Cetin2019}, non-compliance \cite{Maass2021}, or malware infections \cite{Cetin2018}.

Most studies attempted to reach the affected operators via emails to WHOIS or abuse contacts, or common aliases as defined in RFC~2142 \cite{rfc2142}. Some also attempted to work with intermediaries like \glspl{CERT} and vulnerability clearinghouses \cite{Kuhrer2014,Li2016Usenix,Stock2016,Cetin2017}, \glspl{ISP} \cite{Cetin2016,Cetin2019}, or Google \cite{Li2016WWW,Zeng2019,Li2016Usenix}. 
Two studies evaluated more labor-intensive contact channels. Stock \emph{et al.} used manually-collected contact information for a variety of channels like letters, phone calls, and social media, obtaining mixed results \cite{Stock2018}. Maass \emph{et al.} used manually-collected email and postal addresses and found letters to be significantly more effective than emails \cite{Maass2021}.

Many studies reported delivery problems \cite{Stock2016,Stock2018,Cetin2016,Durumeric2014,Li2016Usenix,Cetin2017,Maass2021}, citing high bounce rates \cite{Cetin2017,Stock2016,Cetin2016} and spam filters \cite{Stock2016,Stock2018} as hurdles for message delivery. Even if the messages were delivered, recipients often distrusted these unsolicited mails \cite{Stock2018,Cetin2018,Cetin2019,Zeng2019,Maass2021} or performed extra steps to validate them \cite{Cetin2018,Cetin2019,Maass2021}.

More detailed messages appear to increase remediation rates \cite{Vasek2012,Li2016Usenix,Cetin2016} and trust in the message~\cite{Stock2018}. Recipients also requested assistance in validating that the notified issue was fixed \cite{Zeng2019,Cetin2017,Li2016WWW,Maass2021}. However, one study reported that providing a verification tool did not have a large effect on remediation \cite{Cetin2017}.

\section{Methodology}
\label{sec:methodology}
\begin{figure}
    \includegraphics[width=\linewidth]{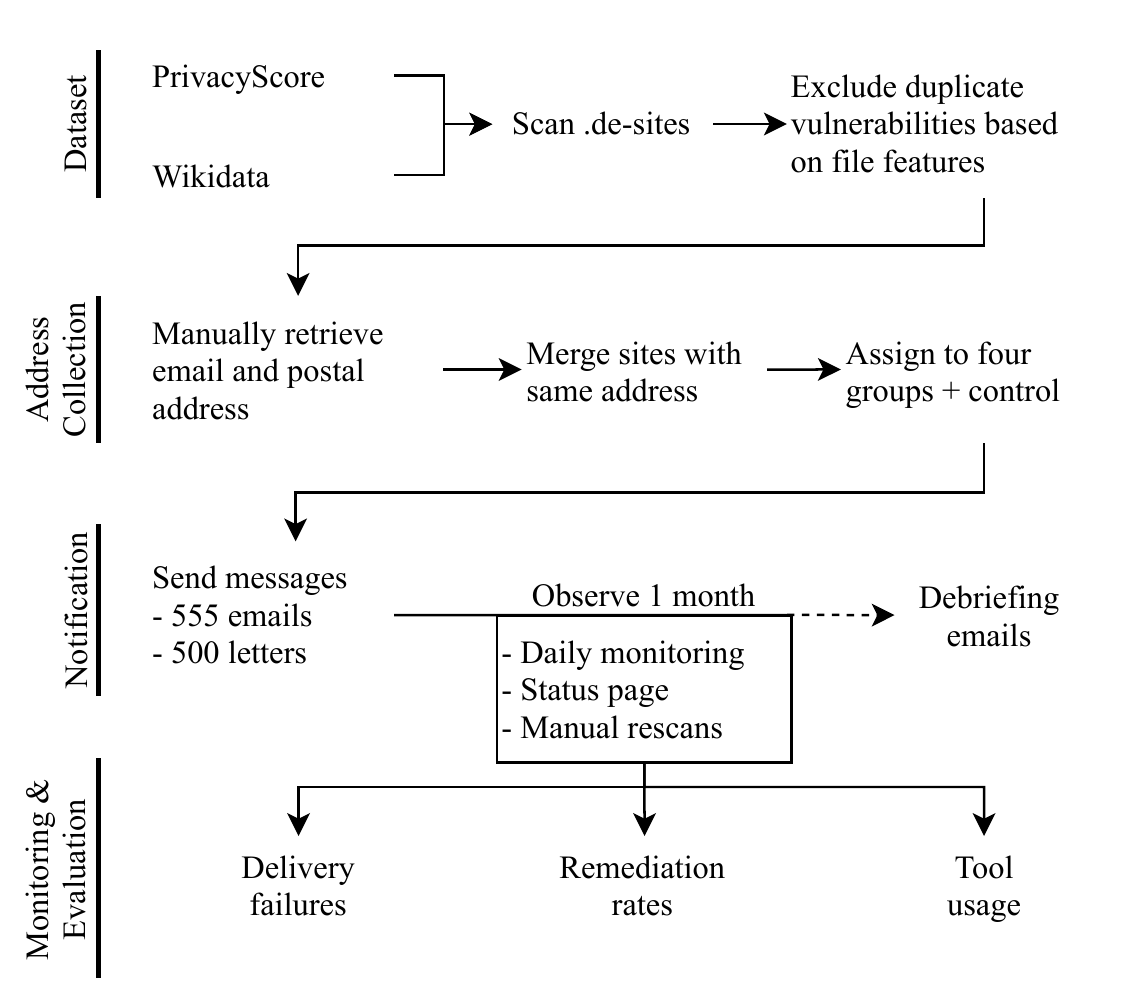}
    \caption{Methodology overview}
    \label{fig:methodology}
    \Description{A flowchart demonstrating the procedure described in the text. The dataset is combined from the PrivacyScore and Wikidata datasets, and .de-Sites are selected and scanned. Duplicate vulnerabilities are detected and excluded. Afterwards, contact information is manually collected for all sites and used to merge sites with the same address. The remaining sites are assigned to four experimental groups, plus the control group. We then send 555 emails and 500 letters and observe the recipients for a month (daily monitoring of their website, use of the self-service status page, and use of manual scans) before sending the debriefing messages. In the end we evalute the delivery rates, remediation rates, and usage of the tool we provided.}
   \end{figure}

In this section we present our vulnerability notification study. We introduce the vulnerabilities we used as a basis for our notification study and their respective detection technique, and the underlying dataset. We discuss the design of our study and partitioning of the test set. Finally we cover our monitoring system and online checking tool that was provided to the recipients. We close by describing our evaluation strategy and discussing the ethics of our research. \autoref{fig:methodology} provides an overview of the process.

\subsection{Vulnerabilities}\label{sec:methodology:vulns}
As a dataset for our vulnerability notification study, we collect a set of websites with different vulnerabilities that expose private information to the public due to misconfiguration of a webserver or unintentional placement of sensitive files in a public directory. 
We briefly describe the different vulnerabilities in the following.

\paragraph{Cryptographic Keys}
Websites are frequently secured using public/private keypairs, either for use in encrypted \gls{TLS} connections, or for authenticated remote access to the server using \gls{SSH}. The security of these schemes relies on keeping the private key secret. However, inexperienced system operators may place these sensitive files in publicly accessible locations on their webserver, e.g. \texttt{example.com/key.pem}.

\paragraph{Database Backups}
Databases can contain sensitive information like customer data, passwords, or even payment information. It is thus imperative to keep them private. System operators frequently perform backups of these systems by serializing the data into files using utilities like \texttt{mysqldump} \cite{mysqldump}. The generated files may accidentally (or intentionally) be placed in publicly-accessible paths on webservers and thus have the potential to leak information to unauthorized parties.

\paragraph{VCS Repositories}
\glspl{VCS} like Git or SVN are used to manage source code files while developing a system. They contain code and, in some cases, private configuration data. On a technical level, they use a hidden folder (\texttt{.git/} or \texttt{.svn/}) where they manage the history of the source code. If such a hidden folder is publicly accessible (as is the default for popular webservers), it can be used to retrieve the source code of the website \cite{GitInternetwache} and thus potentially expose credentials or security vulnerabilities in the code. Prior research by Stock \emph{et al.} also considered this issue and found it to be wide-spread \cite{Stock2018}.

\paragraph{Server Status Information}
Web servers like Apache can be configured to display information about the server and open connections under special URLs like \texttt{example.com/server-info} \cite{serverinfo} or \texttt{example.com/server-status} \cite{serverstatus}. These status pages contain information that are not intended for the public and could be used to infer how much traffic a website is receiving, if it is running outdated software, or other sensitive information like session tokens.

\paragraph{\phpinfo{} Files}
The scripting language PHP is widely used in web development. It contains a special command, \texttt{phpinfo()} \cite{phpinfo}, which will print out information about the PHP version, loaded extensions, and information about the environment and \gls{OS} it is running on. While not directly harmful in itself, this information can be used to check if the server is running outdated software with known vulnerabilities, or leak secret information encoded in environment variables. It is thus advisable not to keep this information publicly available.


\subsection{Dataset}\label{sec:methodology:detection}
We assemble our dataset by scanning a large set of domains with a custom vulnerability scanning system. We then manually collect contact information for all vulnerable sites. We describe the process in more detail in the following.

\paragraph{Data Sources}
The dataset of domains that serves as starting point for the study is composed of two parts. The first part was provided by the \emph{PrivacyScore} project \cite{maass2017privacyscore}, which, amongst several other checks, scans domains for exposed files. The operators provided us with information about approximately 700 exposed files spread accross 600 different domains, from which we include all domains under the \texttt{.de} \gls{TLD} which are still vulnerable at the beginning of our study. This results in 248 vulnerabilities spread across 234 websites.
In order to expand the dataset, we query the \texttt{official website} attribute of Wikidata.org and use the approximately 35\,000 returned \texttt{.de}-domains as the second part of our dataset.
The resulting combined dataset is used as the input to our vulnerability scanner.
We do not use a dataset of popular websites like the Alexa toplist, as these websites are likely to have been notified about vulnerabilities before and thus might be prejudiced about such notifications. On the other hand, being listed in Wikidata.org guarantees a certain level of relevance.
Additionally, toplists are known to have systemic problems \cite{Scheitle2018,Pochat2019}, and alternatives like the Tranco toplist \cite{Pochat2019} did not exist when the study was conducted.

We only consider \texttt{.de}-domains as we want to limit our notification study to a German-speaking population and also rely on the presence of imprints for contact retrieval, which are required by German legislation. We further exclude all German universities from our dataset, as they had already been notified by the PrivacyScore team in a different study \cite{Mueller2018}. Due to the origins of our dataset, most of the targeted websites belong to people and organisations of public interest. 

\paragraph{Detecting Vulnerabilites}
Our custom scanner consecutively requests a set of paths on each website with a \texttt{GET} request and downloads the first 20 kilobytes of the response if the response code is 200. Thereafter it verifies the contents with a regular expression. If it matches, we add the website and its exposed files to our dataset. The requested paths are chosen based on common filenames and in some cases customized to contain the name of the domain (e.g. \texttt{website.com/website.pem}). A full list can be found in \autoref{tab:vulnpaths} in the Appendix. To account for websites being temporarily unavailable, we scan the dataset two times on different days and discard all websites detected as non-vulnerable in both scans. 
For ethical reasons we include a link to a project website within the User-Agent header, which contains a description of our scans and guidance of how to opt out of our study.
Our scans detected 1830 information leaks spread over 1736 different websites.

To account for vulnerabilities with the same source, e.g. hosted on the same server with a common configuration, we extract characteristic features from the exposed files and check whether they are shared with other sites. If this is the case, we exclude all but one from the study, as their remediations are related.
In total, we identify 79 duplicate vulnerabilities that trace back to 23 common sources.

\paragraph{Gathering Contact Information}
For each vulnerable website, we try to retrieve an email address as well as a postal contact address by manually checking the website for an imprint or a contact page.
We prefer technical contacts over general-purpose addresses if both are given.
%
Collecting both the email and the postal address for all domains also allows us to further deduplicate the domains in the dataset by merging related websites (run by the same operator, e.g. a publisher, or a music label that manages separate websites for their artists). In these cases, only a single message is sent that notifies the recipient about all vulnerabilities at once. This measure further reduces the likelihood that a single operator can have an undue impact on the overall results if it is managing several websites. In the following, we use the term \emph{recipient} to describe a single contact (individual or organizational) that controls one or more websites.

\subsection{Notification Groups}\label{sec:methodology:groups}
Our experiment uses two experimental factors: the delivery medium and the presence or absence of a detailed attack scenario description.
All messages also contain a personalized link to a self-service tool (described in \autoref{sec:methodology:tool}) that allows recipients to validate if the vulnerability persists and to trigger a manual scan to validate their remediation attempts. 
The full message texts are shown in \autoref{appx:notification}.

\paragraph{Notification Medium}
Previous research showed that vulnerability notifications by email have only limited success in reaching the recipients \cite{Cetin2017,Stock2016,Cetin2016} and encounter issues such as spam filtering \cite{Stock2016,Stock2018}. We thus evaluate the impact of using an alternative delivery medium, that is, compare emails to physical letters. We denote these notification classes as \mail{} and \letter{}, respectively.

Email messages are sent using a purpose-specific email account linked to our research group (\texttt{web-survey@group.university.de}). The email account is hosted on the Google Apps for Education platform and thus uses the Google Mail infrastructure for message delivery. Emails are sent as plaintext, as Stock \emph{et al.} \cite{Stock2018} previously found HTML emails to be less effective.

Physical letters are sent using the official letterhead of our research group, and contain a scanned signature from one of the researchers. The letterhead also contains contact information for letters, fax, and emails, where it lists the same purpose-specific email account. It does not list a telephone number. See \autoref{appx:notification} for an example letter.

\paragraph{Attack Scenarios}
As the risk posed by some of the vulnerabilities may not be obvious, we compare two different framings for our messages: The \emph{baseline} message simply contains information about the detected information leaks, without discussing the potential impact. The second class of messages, denoted with an \groupplus{} suffix (e.g., \mailplus{}), also contains a description of an \emph{attack} enabled by the vulnerability, under the assumption that such an attack scenario illustrates the risks of the vulnerability and thus serves as an incentive to remediate.

\paragraph{Group Assignment}
Adding an unnotified control group, we have a total of five experimental groups. Before assigning the groups, we scan all vulnerable websites again and remove those that have already been remediated. Recipients are then assigned randomly to the different groups, without considering address availability. Recipients assigned to a medium for which we did not find an address are not contacted and not considered in later parts of the evaluation, thereby slightly reducing the sample size, but avoiding self-selection bias.

As remediation behavior may differ for different vulnerabilities, the different vulnerability classes are stratified between the groups. \autoref{tab:groupsizes} shows the final distribution of vulnerabilities in the groups, considering only recipients for which the correct address type is available (i.e., that were actually notified). Due to the limited number of letters we can send, \letterplus{} has fewer members than the other groups. We will consider the possible effects of this imbalance between the groups in \autoref{sec:discussion:limitations}.


\begin{table}
    \centering
    \caption{Number of notified recipients per group and vulnerability (recipients can be affected by more than one vulnerability)}
    \begin{tabular}{lrrrrr|r}
        \toprule
        Group &  Status &  VCS &  DB &  Key &  \phpinfo{} & Total\\
        \midrule
        \mail{}       & 17 & 18 & 2  & 1 & 243  & 275 \\
        \mailplus{}   & 13 & 16 & 3  & 0 & 253  & 280 \\
        \letter{}     & 17 & 19 & 3  & 1 & 250  & 287 \\
        \letterplus{} & 14 & 18 & 2  & 0 & 180  & 213 \\
        \control{}    & 21 & 18 & 4  & 0 & 269  & 304 \\
        \midrule
        Total:        & 82 & 89 & 15 & 2 & 1196 & 1359 \\
        \bottomrule
    \end{tabular}
    \label{tab:groupsizes}
\end{table}

\paragraph{Experiment Timeline}
After monitoring the websites for five days to be sure the vulnerabilities persisted, we finalize the groups on June 10th, 2018, and send the letters on June 11th. To account for the higher delivery times, we hold off on sending the emails for two days, finally sending them on June 13th. Due to the high effort and cost of sending postal mail we do not send any reminder messages. 
We monitor remediation for one month before finishing the experiment. For ethical reasons, we then (re-)notify all recipients that have not remediated yet by email, including those in the \control{} group, informing them that are (still) vulnerable to give them an opportunity to remediate.

\subsection{Monitoring}\label{sec:methodology:monitoring}
Each night we initiate a check of all websites that we still consider to be vulnerable. If our scan shows a vulnerability as remediated, we repeat the scans for four days to confirm that the reading was not caused by a transient server or scan error. We say that a recipient has remediated when \emph{all} vulnerabilities on \emph{all} of their websites are fixed.


\subsection{Online Checking Tool}\label{sec:methodology:tool}
\begin{figure}
    \centering
 \includegraphics[width=\linewidth]{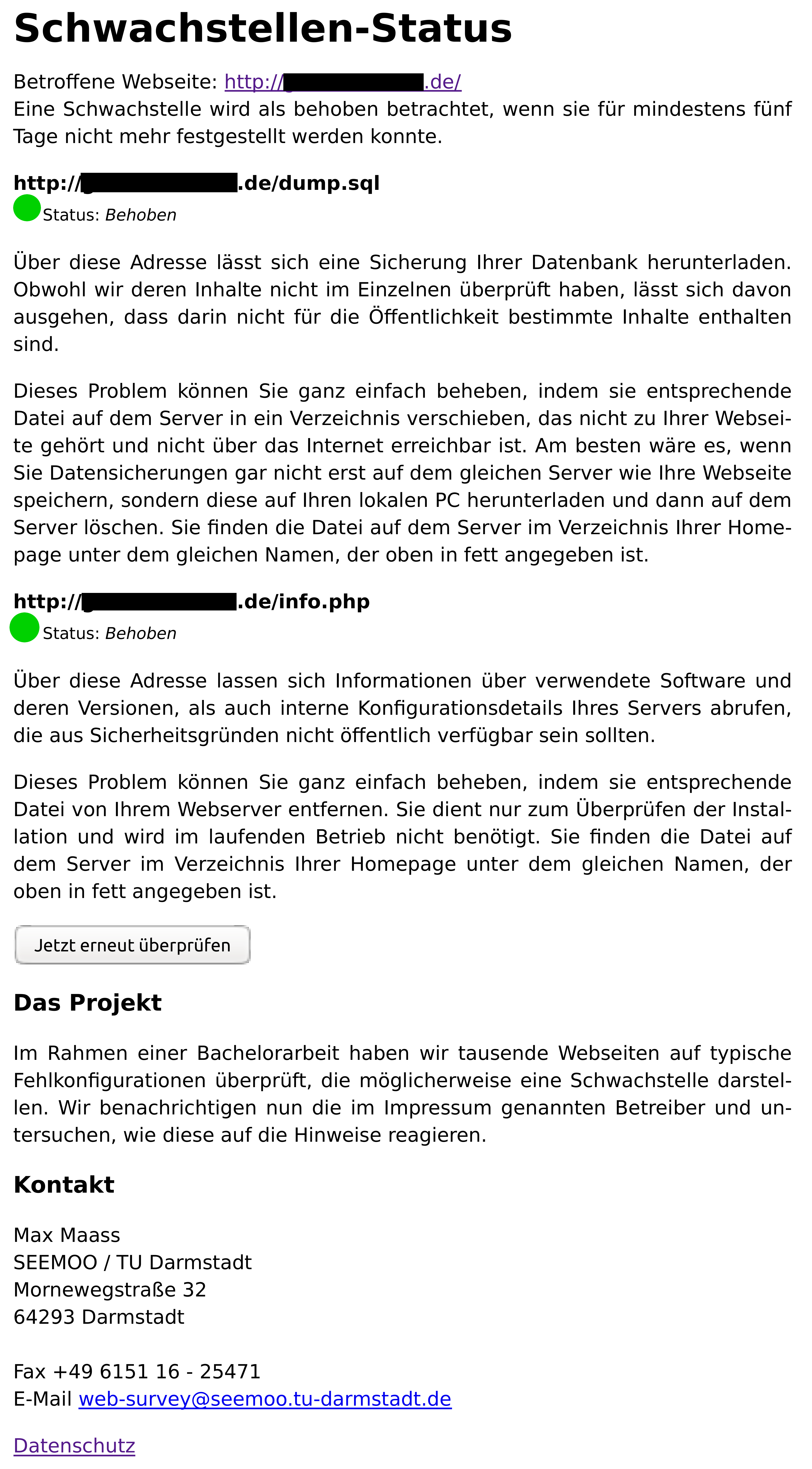}
 \caption{The German-language status page}
 \label{fig:check-tool}
 \Description{A German-language version of the status page, showing the status and explanations of two vulnerabilities. It also contains a button to trigger a rescan of the vulnerabilities, an explanation about the project, and contact information of the researchers.}
\end{figure}

In order to ease remediation for the recipients of our vulnerability notifications, we provide an online status page that can be accessed with a personalized link included within the notification (cf. \autoref{fig:check-tool}). It lists the vulnerabilities, each with the current remediation status indicated by a simple traffic light scheme. A \emph{red} dot indicates the vulnerability has been detected by the last daily scan, while a \emph{yellow} dot indicates the last check was negative, but we have seen the vulnerability within the last five days. After five consecutive negative checks the dot turns \emph{green}. Each vulnerability is accompanied with information about its impact and guidance on how to fix the respective issue. 

To further help the recipients to verify the success of the remediation, the status page offers the possibility to manually trigger a check every 15 minutes, which will immediately switch the color of the respective vulnerability from \emph{red} to \emph{yellow} if it has been fixed. Recipients are only able to scan their own website(s), as identified by the token included in their personalized link. In order to evaluate the adoption of this status page, we record timestamps of the page visits and manually triggered scans.

\subsection{Evaluation}\label{sec:methodology:eval}
\begin{table*}
    \centering
    \caption{Reachability of the recipients per contact group}
    \begin{tabular}{lrrrrr}
        \toprule
        Group & Assigned & No Contact & Bounced & Reached & Unknown \\
        \midrule
        \mail       & 302 & 27\ \ \ (8.94\,\%) & 4\ \ \ (1.32\,\%) & 76\ \ \ (25.17\,\%) & 195\ \ \ (64.60\,\%)\\
        \mailplus   & 302 & 22\ \ \ (7.28\,\%) & 6\ \ \ (1.99\,\%) & 74\ \ \ (24.50\,\%) & 200\ \ \ (66.23\,\%) \\
        \letter     & 304 & 17\ \ \ (5.61\,\%) & 3\ \ \ (0.99\,\%) & 97\ \ \ (32.01\,\%) & 187\ \ \ (61.51\,\%) \\
        \letterplus & 224 & 11\ \ \ (4.89\,\%) & 3\ \ \ (1.34\,\%) & 58\ \ \ (25.78\,\%) & 152\ \ \ (67.86\,\%) \\
        \midrule
        Sum         & 1132 & 77\ \ \ (6.80\,\%) & 16\ \ \ (1.41\,\%) & 305\ \ \ (26.94\,\%) & 734\ \ \ (64.84\,\%) \\
        \bottomrule
    \end{tabular}
    \label{tab:groupswithcontacts}
\end{table*}

To evaluate the effectiveness of our notifications, we measure the remediation rates over time. Only notified recipients are considered (i.e., those that were assigned to a medium for which no contact information was available are not counted into the total number). Not all email servers send a notice if they discard a message as spam, leading to an unknown number of silently discarded messages for the \mail{} and \mailplus{} groups. To avoid introducing any biases in comparison to the letters, we do not attempt to exclude recipients where message delivery failed from the evaluation, thereby slightly lowering reported remediation rates compared to studies that exclude these messages.

As previously described, a website counts as remediated if all of its vulnerabilities are remediated, and a recipient counts as having remediated if all of their websites are remediated.
Considering recipients instead of websites ensures that all recipients make the same contribution to the overall remediation rates, regardless of the number of websites they control.

However, such an evaluation will only give us information about how \emph{our sample} behaved. As some combinations of factors result in very small samples sizes, we want to estimate how much variation we could expect, if we were to repeat the experiment with another dataset of websites with the same characteristics. For this, we turn to \emph{bootstrapping}. Bootstrapping allows us to approximate the variation in the results we would expect if we were to repeat the experiment many times with similar samples of websites. It can thus serve to quantify how much uncertainty remains in our results.

Mathematically speaking, our evaluation considers the empirical distribution of remediation vs. non-remediation, $F^*$, for our sample of $n$ recipients, $x_1,\dots{},x_n$, drawn from the base distribution $F$ (i.e., the distribution we would have obtained, had we performed a notification experiment with \emph{all} affected recipients in Germany). We can use this sample to estimate the \emph{variation} of a statistic $u$ computed over $F$. For this, we take a sample with replacement of size $n$ from the (known) \emph{empirical} distribution $F^*$, denoted $x^*=x^*_1,\dots{},x^*_n$, and compute a statistic $u^*$ over it. According to the \emph{bootstrap principle} \cite{efron1979bootstrap} (as described in \cite{orloff2014bootstrap}), the variation of $u^*$ approximates the variation of $u$ well. We can thus compute many resamples $x^*$, compute $u^*$ for each of them, and then compute the measure of choice for the variation from these results.

While this technique has some limitations (in particular, the implied assumption that the original sample was representative for $F$, and that $x_1,\dots{},x_n$ are independent), it can serve to give an indication of how much we would expect the statistic to vary. We use this technique with 10\,000 iterations to compute the 1st and 3rd quartile in addition to the median of the remediation rates on each day. While this cannot repair issues caused by very small samples sizes, it serves to indicate how imprecise we can expect our results to be due to them.

\begin{figure*}
\centering
    \subfloat[Full Groups]{{\includegraphics[width=.45\linewidth]{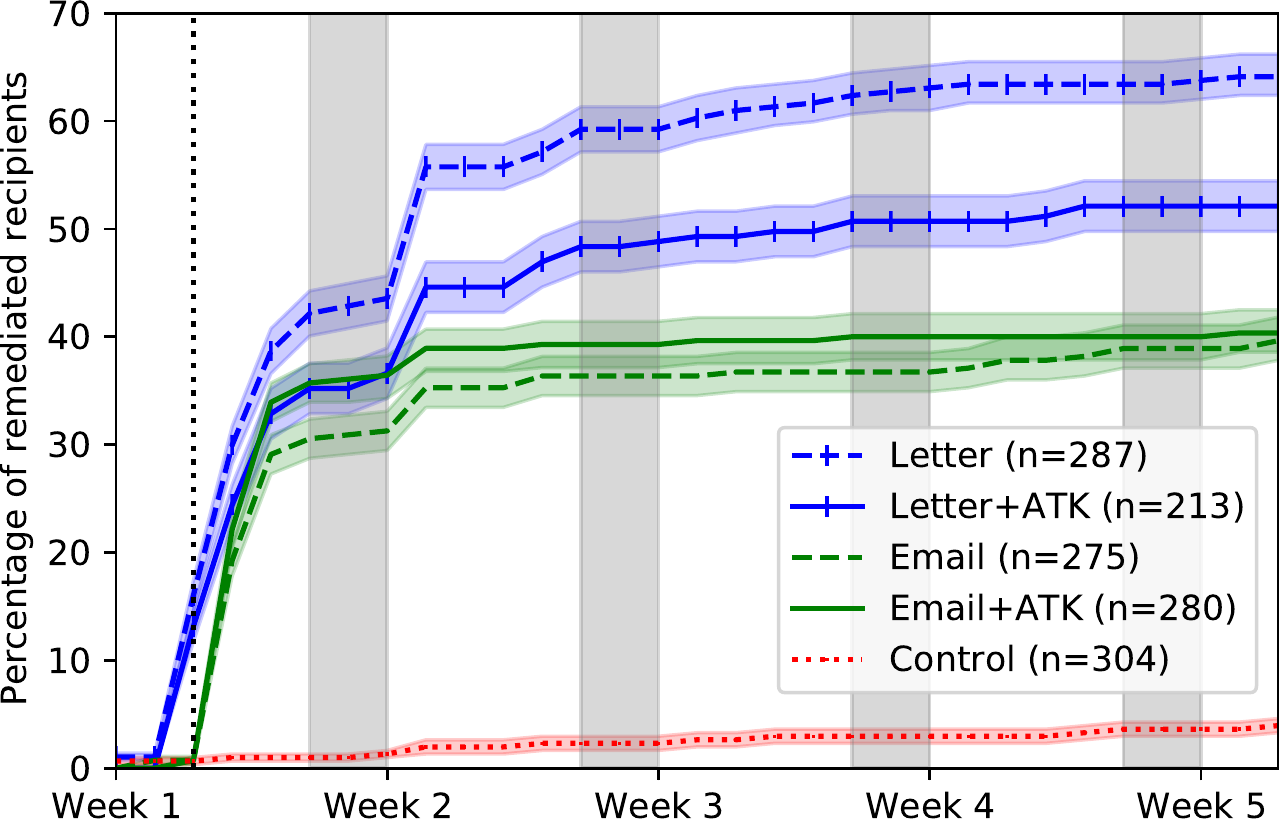}}}%
    \qquad
    \subfloat[Vulnerabilities (Cryptographic Keys omitted for readability)]{{\includegraphics[width=.45\linewidth]{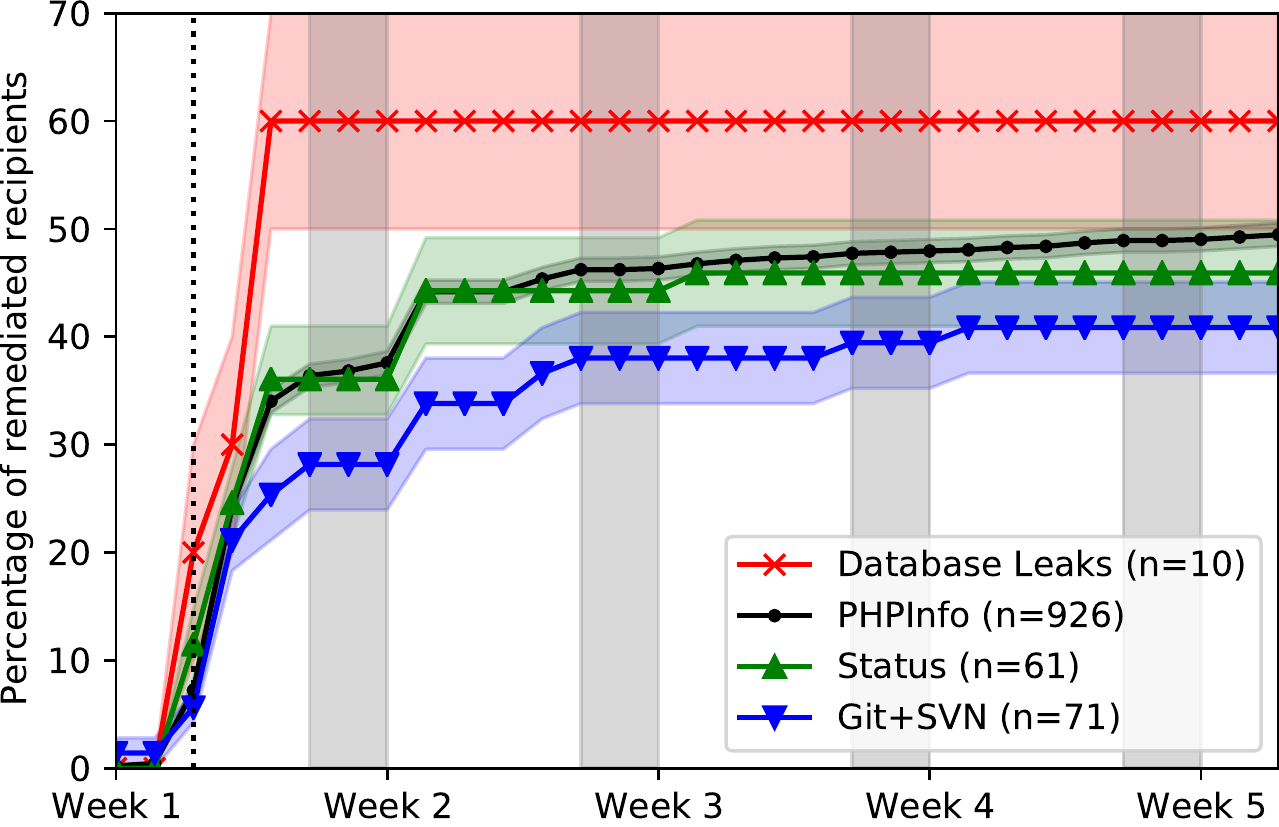}}}\\
    \subfloat[Medium]{{\includegraphics[width=.45\linewidth]{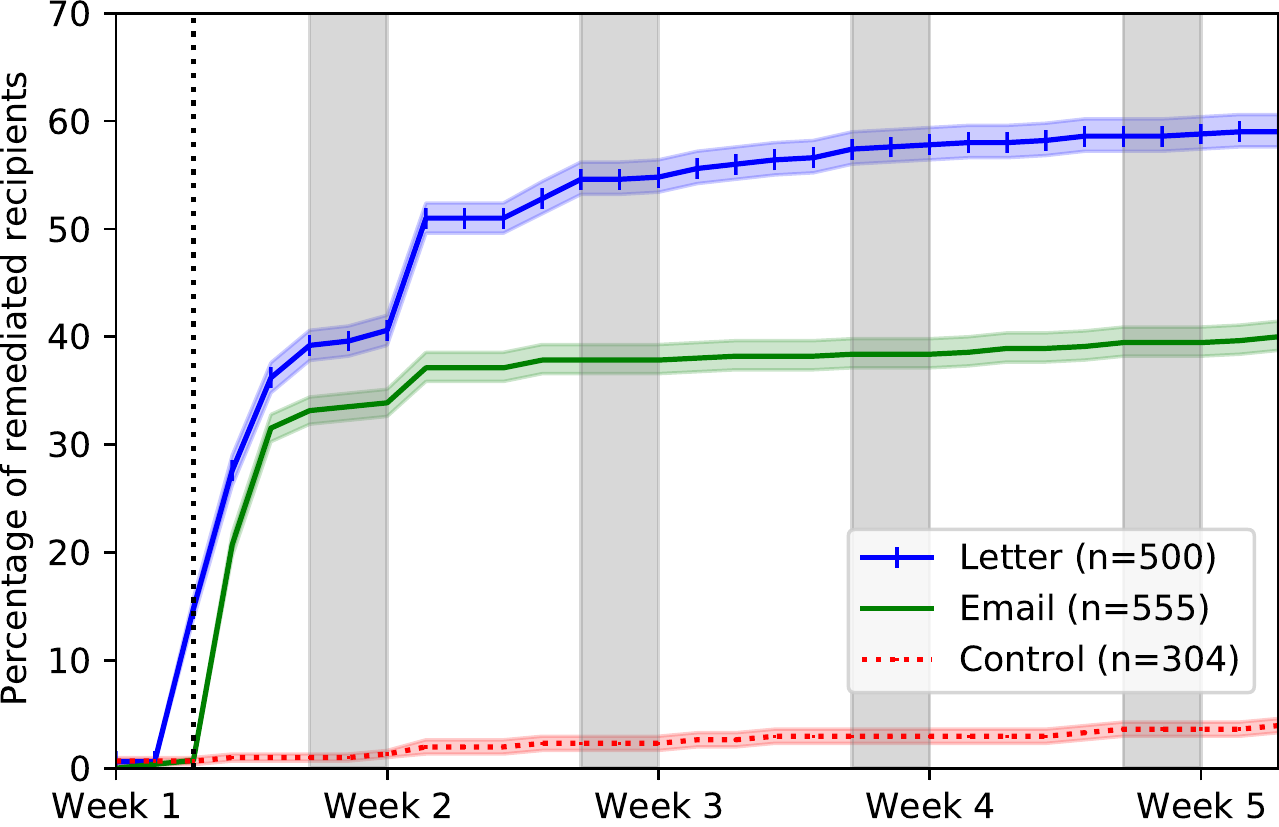}}}%
    \qquad
    \subfloat[Scenario]{{\includegraphics[width=.45\linewidth]{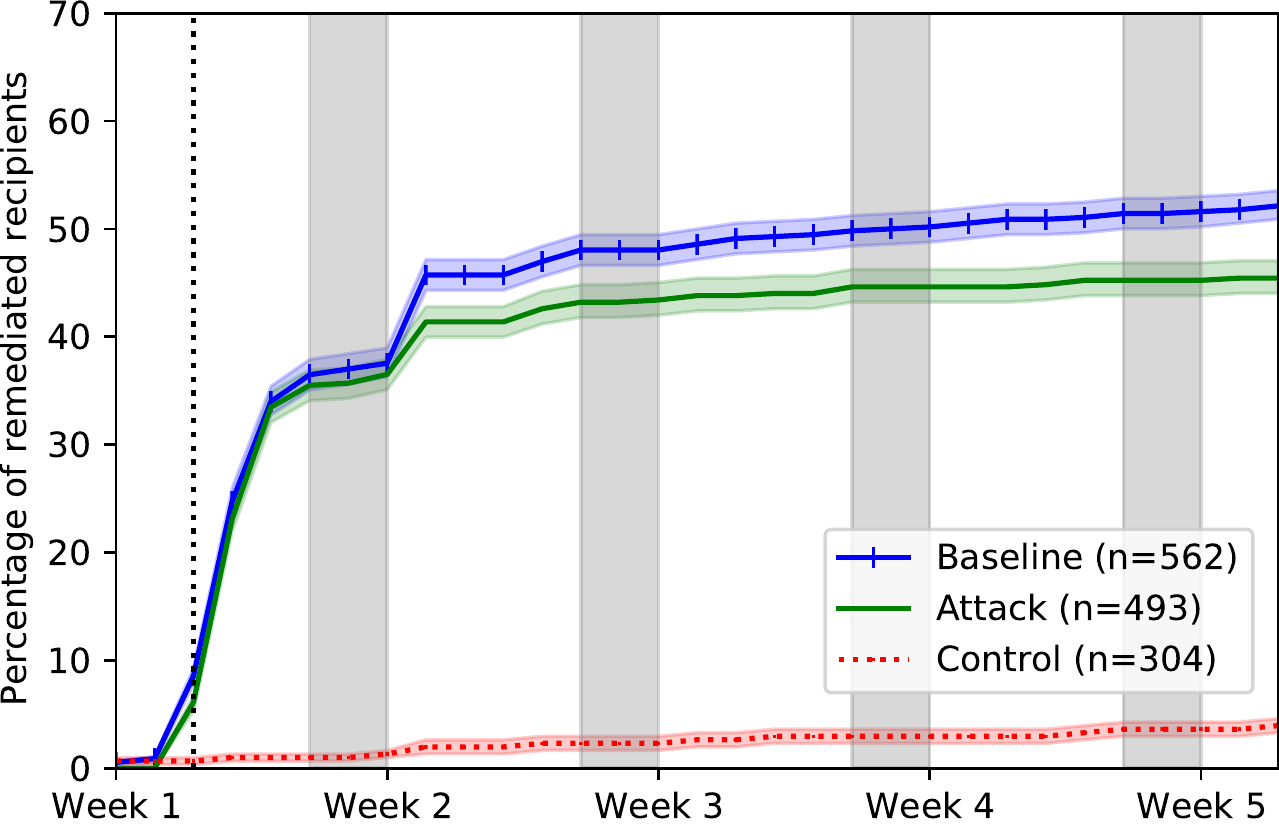}}}
    \caption{Median, 1st and 3rd quartiles for the remediation rates in different experimental groups, vulnerabilities, mediums, and presence of attack scenarios. Gray areas denote weekends, dotted line shows when the emails were sent.}
    \label{fig:groupcomparison}%
    \Description{A set of four figures. The top left figure shows the remediation rates for the experimental groups (Mail, Mail+ATK, Letter, Letter+ATK) over the month of the experiment. The final values are given in Table 3. Lines rise quickly for the first week and then reach a near-linear growth for the remaining weeks. The top right figure shows the different vulnerability types. They mostly show similar trends, but many show large distances between the first and third quartile due to their low sample sizes. Again, the numbers are given in the table. The bottom left figure shows the differences between the mediums (letter and email). They start growing at approximately equal rates, but the letters end up growing further than the emails. The last figure shows the same for the presence of the attack scenario. Overall, the baseline scenario exceeds the attack scenario by approximately 6 percentage points.}
\end{figure*}

\subsection{Ethical Considerations}\label{sec:methodology:ethics}
Large-scale vulnerability scanning operates in a legal and ethical gray area. We avoid collecting sensitive information with our scans by limiting the amount of data we download, and discarding the data after the end of the study. Through this, we also avoid putting any undue strain on the infrastructure of the site operator. Our scanner identifies itself with a custom user agent and a reverse DNS entry for the IP of the scanning machine, and offers website operators a way to opt out of the study. The sent messages state that they are part of a study, and contain our contact information to allow recipients to opt out as well. All unnotified recipients for whom contact information is available are notified by email after the end of the study to give them an opportunity to remediate.

At the time of the study, our institution did not require ethics approval for this type of research. While we thus did not seek out an ethical review for this study, we successfully obtained approval for a different study with a substantially similar setup that employed similar safeguards. We also discussed the study with legal experts to ensure its legality in our jurisdiction.

\section{Results}
\label{sec:results}
In this section, we investigate the effect of our notifications, the use of our self-service check tool, and briefly discuss the interactions with the recipients. These results will be interpreted in more detail in \autoref{sec:discussion}.

%
%
%
%

\subsection{Notifications}
\label{sec:results:notifications}
\autoref{tab:groupswithcontacts} gives an overview of the delivery success of the notifications. In total, 77 (6.8\,\% of the original 1132 recipients, not counting the 304 in the control group) were assigned to a medium for which no address could be found, and thus were not notified. 10 emails (1.7\,\%) and 6 letters (1.1\,\%) could not be delivered and were returned to the sender. The true number of undelivered emails may be higher, as spam filters may have silently discarded messages. At least 305 messages (26.9\,\%) were read (i.e., we received a non-automated response or the self-service tool was accessed), which leaves 734 messages (64.8\,\%) in an unknown state.

\paragraph{Overall Remediation Rates}
\autoref{tab:remediationrates} shows the median remediation rates for the different groups, and their first and third quartiles. Overall, between \mailmedian{} (\mail{}) and \lettermedian{}\,\% (\letter{}) of recipients remediated all issues with their website within the month of the study, depending on the notification channel and presence of the attack scenarios. All groups outperformed the control group by a large margin, which had a remediation rate of only \controlmedian{}\,\%.
\autoref{fig:groupcomparison}a shows that most remediations take place within two weeks, after which only comparatively few additional vulnerabilities are fixed.

\paragraph{Communication Medium} 
Overall, letters markedly outperformed the email group (cf. \autoref{fig:groupcomparison}c), showing remediation rates of \alllettermedian{}\,\% compared to the \allmailmedian{}\,\% achieved by sending emails.
After a week, almost half the letter recipients had already remediated, compared to only a third of those notified by email.

\paragraph{Attack Scenarios}
The impact of including attack scenarios is less pronounced and, interestingly, different across mediums---while the attack scenario slightly increased remediation for emails, it actually greatly \emph{decreased} remediation for letters (cf. \autoref{fig:groupcomparison}a), reducing the remediation rate by 12.4 percentage points. 
When considering vulnerabilities separately (cf. \autoref{tab:vulnsinframings}), we see that this effect is dominated by the \phpinfo{} group, which showed a large drop in remediation rates, loosing 8.3 percentage points when including descriptions of attacks.
Interestingly, this effect is not spread evenly: the \mailplus{} and \mail{} groups have almost identical performance for the \phpinfo{} vulnerability (40.7 vs. 41.1\,\%), while \letterplus{} has a much worse performance than \letter{}, dropping from 65.2 to 51.1\,\% remediation at the end of the study timeframe.
The other vulnerability classes actually saw increased remediation rates when adding the attack scenarios (between 6.3 and 10.7 percentage points), although the small sample sizes lead to a large spread in the quartiles, which overlap heavily.
In aggregate, this leads to messages featuring the attack scenario achieving a remediation rate of \allplusmedian{}\,\%, and being outperformed by their less explicit counterparts with \allplainmedian{}\,\% remediation (cf. \autoref{fig:groupcomparison}d).

\paragraph{Vulnerability Type}
\autoref{fig:groupcomparison}b shows the remediation rates of the different vulnerability types, considering only those not assigned to the control group. Database leaks show the highest remediation rates with \databasemedian{}\,\% ($n=10$), however, their low overall number leads to limited expressiveness. 
\phpinfo{} leaks were fixed by \phpinfomedian\,\% of recipients ($n=926$), with the server status ($n=61$) and \gls{VCS} ($n=71$) misconfigurations following with \statusmedian{} and \vcsmedian{}\,\%, respectively.
The figure omits the publicly available cryptographic keys ($n=2$) for readability---one of them was removed by the end of the study, while the other remained available, leading to a remediation rate of 50\,\%.

\paragraph{Effect of Reachability}
The obtained results, especially for the message medium, raise the question if we are only measuring how well messages are delivered, or if the medium also has an effect outside of the rates of successful delivery.
We thus repeat the evaluation with only those recipients that either used our self-service tool or from whom we received a non-automated response.
Both of these indicate that the message was read.
By its nature, such a sample is heavily self-selected and unrepresentative, but it may serve as an indication what factors influence the most motivated recipients (i.e., those that either contact us or use our tools).

In this sample of 305 recipients, we observe overall high remediation rates of 85.3\,\% for emails and 90.3\,\% for letters (cf. \autoref{fig:groupcomparison_reached} in the Appendix).
Both versions of the messages (baseline and attack) achieve almost identical average performance (88.4 vs 87.1\,\%, respectively), but when considering all four combinations of medium and attack scenario, we still see that the attack scenario seems to help emails and hurt letters, although the gap has shrunk.
The trends for the different vulnerability classes remain similar as well.

\subsection{Self-service Tool}
\label{sec:results:tool}
\begin{table}[t]
    \centering
    \caption{Median, 1st (Q1) and 3rd (Q3) quartiles of bootstrapped remediation rates in percent for different groups at the end of the study timeframe}
    \begin{tabular}{lrrrr}
        \toprule
        Group        & n   & Median            &  Q1  &  Q3  \\
        \midrule
        \mail        & 275 & \mailmedian       & 37.5 & 41.5 \\
        \mailplus    & 280 & \mailplusmedian   & 38.2 & 42.1 \\
        \letter      & 287 & \lettermedian     & 62.6 & 66.1 \\
        \letterplus  & 213 & \letterplusmedian & 49.5 & 54.2 \\
        \midrule
        All emails   & 555 & \allmailmedian    & 38.4 & 41.3 \\
        All letters  & 500 & \alllettermedian  & 57.4 & 60.4 \\
        \midrule
        All baseline & 562 & \allplainmedian   & 50.8 & 53.5 \\
        All \groupplus{} & 493 & \allplusmedian    & 43.7 & 46.8 \\
        \midrule
        \phpinfo     & 926 & \phpinfomedian    & 48.3 & 50.5 \\
        \gls{VCS}    & 71  & \vcsmedian        & 36.6 & 45.1 \\
        Status       & 61  & \statusmedian     & 41.0 & 50.8 \\
        Database     & 10  & \databasemedian   & 50.0 & 70.0 \\
        Keyfile      & 2   & 50.0              & 50.0 & 50.0 \\
        \midrule
        \control     & 304 & \controlmedian    & 3.6  & 4.9  \\
        \bottomrule
    \end{tabular}
    \label{tab:remediationrates}
\end{table}

\begin{table}[t]
    \centering
    \caption{Median and quartiles of bootstrapped remediation rates for different vulnerability types at the end of the study timeframe, with and without attack scenarios. (Database and Keyfile omitted due to low sample size)}
    \begin{tabular}{llr|rrr}
        \toprule
        & Group & n & Median & Q1 & Q3 \\
        \midrule
        \parbox[t]{2mm}{\multirow{3}{*}{\rotatebox[origin=c]{90}{Baseline}}} & \phpinfo & 493 & 53.3 & 51.9 & 55.0 \\
        & Status & 34 & 41.2 & 35.3 & 47.1 \\
        & VCS & 37 & 37.8 & 32.4 & 43.2 \\
        \midrule
        \parbox[t]{2mm}{\multirow{3}{*}{\rotatebox[origin=c]{90}{Attack}}} & \phpinfo & 433 & 45.0 & 43.4 & 46.7 \\
        & Status & 27 & 51.9 & 44.4 & 59.3 \\
        & VCS & 34 & 44.1 & 38.2 & 50.0 \\
        \bottomrule
    \end{tabular}
    \label{tab:vulnsinframings}
\end{table}

\begin{table}[t]
    \centering
    \caption{Percentage of recipients who viewed and used the tool before ($B$) and after ($A$) remediation}
    \begin{tabular}{l|rr|rr}
        \toprule
        Group & $View_B$ & $View_A$ & $Use_B$ & $Use_A$ \\
        \midrule
        \mail       & 23.6\,\%  & 13.5\,\%  & 13.5\,\%  & 7.3\,\% \\
        \mailplus   & 22.1\,\%  & 15.4\,\%  & 15.0\,\%  & 11.4\,\% \\
        \letter     & 25.2\,\%  & 22.0\,\%  & 18.5\,\%  & 15.0\,\% \\
        \letterplus & 22.0\,\%  & 15.4\,\%  & 15.9\,\%  & 9.8\,\% \\
        \bottomrule
    \end{tabular}
    \label{tab:tooluse}
\end{table}

\begin{table}[t]
    \centering
    \caption{Number of contacted recipients and non-automated responses by group and medium}
    \begin{tabular}{lr|rrrr|r}
        \toprule
        Group & n & Email & Phone & Fax & Letter & Sum\\
        \midrule
        \mail & 275 & 29 & 3 & 0 & 0 & 31\\
        \mailplus & 280 & 28 & 2 & 0 & 0 & 30\\
        \letter & 287 & 25 & 3 & 1 & 0 & 29\\
        \letterplus & 213 & 13 & 1 & 0 & 1 & 15 \\
        \midrule
        All & 1055 & 95 & 9 & 1 & 1 & 105 \\
        \bottomrule
    \end{tabular}
    \label{tab:responses}
\end{table}

The self-service tool was accessed by 266 (25.2\,\%) recipients, with 192 (18.2\,\%) performing a manual scan. The distribution over the experimental groups is shown in \autoref{tab:tooluse}. 65.8\,\% of visited status pages were accessed only on one day, although some were viewed on up to 15 separate days (median: 1, Q1: 1, Q3: 2). Similarly, while 37\,\% of tool users required only a single scan, some triggered up to nine scans (median: 2, Q1: 1, Q3: 3). 116 recipients (11\,\% of the total and 60.4\,\% of scan users) scanned their website after remediating to validate that their remediation attempt was successful.

Use of the tool and remediation seem to be linked---90.2\,\% of recipients that opened the tool and 95.8\,\% of those that triggered a scan before attempting a remediation successfully remediated the issue(s) afterwards. We note that this does not imply that the tool \emph{caused} the remediation, as the users were self-selected (recipients that clicked the link clearly received the notification and trusted it enough to open a link, which makes them much more likely to remediate, even without the tool). We thus cannot make any statements about the effect of the tool on remediation.

\subsection{Communication with Recipients}
\label{sec:results:communication}
We received responses from 105 out of the 1055 contacted recipients, not counting bounces and autoreplies (a detailed overview is given in \autoref{tab:responses}).
Most of the respondents were grateful, only two were hostile, interpreting our messages as either unsolicited advertising or fraud. To these, we sent clarifications and offered not to contact them again.
86 respondents stated that the problem had been remediated or that the responsible person had been instructed to fix it. Three of these still had unremediated \phpinfo{} issues at the end of the study timeframe.
Some also explicitly mentioned using the tool we provided and finding it helpful.
We did not receive any opt out requests, only one unspecific abuse notification directed at our network provider.

The majority of respondents sent emails, but we also received one letter and one fax. Interestingly, nine recipients chose to contact us via phone calls. As we did not provide a phone number in the notification messages, they checked the website of the university and research group to find phone numbers, with one person calling the central switchboard of the university and being forwarded via multiple intermediaries until they reached the responsible researcher. We thus consider these numbers to be a lower bound, as some may not have been able to reach the right person in the end.
They often stated that they mistrusted the message and wanted to verify its authenticity using a different channel.

Some recipients asked if we were aware of other issues with their website, or if we could scan additional websites under their control. These, we referred to the scanning service PrivacyScore.org \cite{maass2017privacyscore}, which performs similar checks for information leakage.

\section{Discussion}
\label{sec:discussion}
In this section, we review and interpret our results, discuss the limitations of our study, and identify areas for future research into effective notifications.

\subsection{Effectiveness of Notifications}
\label{sec:discussion:takeaways}
Overall, \allnotifiedmedian{}\,\% of notified recipients remediated, compared to \controlmedian{}\,\% of the control group. 
This demonstrates that our notifications were effective at increasing remediation. However, the different experimental groups show a large spread of remediation rates, ranging from \mailmedian{} to \lettermedian{}\,\%.
This indicates that the different factors of the notification can have a large impact on remediation. We discuss these factors in more detail here.

\paragraph{Manual Address Collection Improves Deliverability}
Previous studies attempting to contact website owners directly \cite{Cetin2016,Stock2018,Stock2016,Zeng2019,Maass2021} frequently struggled with delivery problems, with many studies showing bounce rates of over 10\,\% for WHOIS contacts \cite{Stock2016,Stock2018} (although Zeng \emph{et al.} reported only 3\,\% \cite{Zeng2019}) and over 50\,\% for standard aliases like abuse@ \cite{Stock2016,Cetin2016}.

Two prior studies used manual address collection to overcome this problem.
Stock \emph{et al.} reported no email bounces, but a bounce rate of 26.8\,\% from their letters \cite{Stock2018}. The latter may be related to the international nature of their study, while our study was limited to Germany, where system operators are legally required to disclose a functional postal address.
However, Maass \emph{et al.} reported bounce rates of 3.5 and 5.8\,\% for letters and emails in a similar study of German website operators \cite{Maass2021}, exceeding the 1.1 and 1.7\,\% observed in this study.
This indicates that other factors also influence the delivery success.

Overall, the labor required for manual address collection may be difficult to scale to very large notification campaigns, although it could be justified for smaller or important notifications. Long-term, increased adoption of standardized ways of providing contact information for security notifications, like the proposed security.txt standard \cite{draft-foudil-securitytxt-09}, is needed to facilitate more reliable automated notifications.

\paragraph{Letters are Effective}
Letters provided a large boost in remediation compared to emails, with an increase of almost 20 percentage points. This may be related to a higher \emph{a priori} trust into postal messages, as (at least in Germany) this communication channel is less often abused for spam and scam messages compared to emails. It may also be partially related to the more reliable delivery and lack of spam filters in the postal system. However, even when considering only recipients that reacted to our message, letters still show higher remediation rates than emails, indicating that at least some of their increased effectiveness cannot be attributed to the higher delivery success.

Once again, this increased remediation rate comes at a cost---sending 500 letters cost around 400\,€, and around five hours for the printing and manual enveloping of the messages (although the latter could be avoided through the use of commercial mailing services or machines). However, as with the manual collection of contact information, the added expense could be justified for critical notifications. 

Our results are in agreement with those by Maass \emph{et al.}, who observed an increase in remediation rate of 11.2 percentage points when switching from emails to letters \cite{Maass2021}. At first glance, both results seem to conflict with prior results by Stock \emph{et al.}, who reported only a slight increase in remediation rates for letters compared to fully-automated email notifications \cite{Stock2018}. However, these numbers should not be compared directly, as their dataset contained only operators that did not react to an initial automated message.

\paragraph{Verifiability Fosters Trust}
Gaining the trust of the recipients and convincing them that the message is legitimate is an important challenge when sending unsolicited notifications. While we did not specifically ask about it, several recipients mentioned that recognizing the name of the sending university helped overcome their inherent distrust.
This is in line with previous results by Maass \emph{et al.}, who reported similar results \cite{Maass2021}.
However, some recipients still wanted to ensure that the message was authentic (and not just printed on an official-looking letterhead). These recipients invested considerable effort in searching for a phone number and calling the senders. 
This matches previous studies that reported recipients reaching out to verify the authenticity of the messages \cite{Maass2021,Cetin2018,Cetin2019}.

Similarly, providing a tool for recipients to verify the claims made in the message can help to convince the recipients that the message is correct, especially for non-technical recipients. While previous research found that providing a tool did not significantly increase remediation rates \cite{Cetin2017}, a well-designed tool with documentation about the issue and how to verify and remediate it could increase trust and decrease support requests. Previous studies reported recipients requesting \cite{Zeng2019,Cetin2017,Li2016WWW} and using \cite{Maass2021} such tools, which we also observed in our study.

\paragraph{Tangible Explanations have Limited Impact} 
We saw that adding illustrative attack scenarios seems to have had almost no effect on remediation for the \mail{} groups, and have actively reduced remediation rates for the \letterplus{} group compared to the \letter{} group for the \phpinfo{} vulnerability. This result is surprising, as previous research has shown that more comprehensive messages usually increase remediation rates \cite{Vasek2012,Li2016Usenix,Cetin2016} and trust \cite{Stock2018}. 

We can only speculate about the reasons. 
It may be that the recipients perceived the attack scenarios as an attempt to pressure them into action, which may evoke associations with spam or scam messages. This would have lowered their trust in the message, and thus their willingness to act upon it.
However, it is intuitively unclear why this should only be the case for letters, and not for emails.

It may also be that the expanded explanations actually \emph{detracted} from the perceived urgency, as they provided a more nuanced view of the risk, instead of a blanket statement that the data should not be accessible for security reasons (cf. \autoref{appx:notification}).
This idea is supported by the fact that only the \phpinfo{} group saw decreased remediation rates, while other groups saw increases (cf. \autoref{tab:vulnsinframings}).
However, due to the overall low number of samples in these groups and commensurate large quartile ranges, the statistical basis for claims about their effectiveness is weak.
And again, it is unclear why this would only affect the letters, but not the emails that used identical wordings.

Finally, it may also be related to recipients doing their own research:
when performing a web search for ``phpinfo dangerous'', some results claim that exposing the information is discouraged, but in many cases not actually dangerous.
However, it is unclear why the rate at which recipients seek out further information would \emph{increase} when they are provided with more details in the notification message, and neither is it clear why this should occur for letters, but not for emails.

The only way any of these theories may be plausible is that emails are read by a different group of people than letters, and that these groups interpret the messages differently.
As most organizations only give purpose-specific email addresses for technical matters, but no dedicated postal address, this may have led to a different composition of recipients in the email group compared to the letters.
This could explain the observed differences in behavior between letter and email recipients, although it is impossible to conclusively prove a connection as we did not note if the collected address was technical or general-purpose, and thus cannot differentiate these classes in the evaluation.
It is also unclear why this difference should only affect the \phpinfo{} group and not the others.
Regardless of the exact mechanism at play here, the variation in the results shows that the wording of notifications needs to be carefully considered, and that the optimal wording may also depend on the message medium.

%

\subsection{Limitations}
\label{sec:discussion:limitations}
A large fraction of our dataset consists of \phpinfo{} leaks, which pose a less obvious danger than the other vulnerabilities and are very easy to remediate. However, their remediation rate is only marginally higher than that of other vulnerabilities, thus the impact of this should be limited.
\letterplus{} contains a smaller percentage of \phpinfo{} vulnerabilities than the other groups (83.7 vs. 86.2 to 88.7\,\%), which may affect its overall remediation rates. However, due to the limited difference in remediation rates, the impact should be small and does not explain the large observed differences between \letterplus{} and \letter{}.

We did not attempt to hide that the messages were sent as part of a study. Thus, the study may suffer from observer effects, where recipients behave differently because they are aware that they are part of a study.
Our dataset is geographically limited to German sites, which may introduce a bias if German site operators are in some way different from those in other countries, and increases the effect of sender name recognition (although the actual effect of name recognition is disputed in the literature \cite{Cetin2016,Stock2018,Zeng2019,Maass2021}). It also increases the availability of contact information through the imprint, as providing this information is mandatory in Germany.
Finally, the study was conducted three weeks after the \gls{GDPR} came into force. Thus, the timing of the study may have coincided with a generally increased interest in data protection, the effects of which we are unable to quantify.

\subsection{Future Work}
\label{sec:discussion:future}
In our study, we have shown that letters can help increase remediation rates. This raises the question of which other channels may prove helpful. As previously discussed, Stock \emph{et al.} performed a small ($N=364$) evaluation of channels such as contact forms, phone numbers, letters, and social media \cite{Stock2018}, and found the increase in remediation rates to be too low in relation to the cost to make it worthwhile. A study by Maass \emph{et al.} disagreed and reported letters to significantly increase remediation rates \cite{Maass2021}. Further research is needed into if and when alternative contact channels can be an effective tool for notifications.

In our experiment, we made use of a legal requirement to provide contact information on websites. However, notifications could also make use of other forms of legal requirements, namely, requirements to remediate issues with relevance to data protection or cybersecurity legislation. This may serve as an incentive for remediation. Maass \emph{et al.} found promising results with such an approach \cite{Maass2021}, and Diop \emph{et al.} evaluated different legal environments for relevant legislation \cite{diop_coerce_2019}, which may serve as a starting point for further effort.

The differences in effectiveness between the different message contents also highlight the importance of understanding how recipients perceive notification messages, and which factors influence the trust and perceived urgency. Even though prior studies often reported low engagement with feedback mechanisms \cite{Durumeric2014,Li2016Usenix,Stock2018,Zeng2019}, future studies should include questionnaires or other methods to collect such data, and consider collaborating with researchers from the relevant fields.

If the costs of finding addresses and sending letters is prohibitive, and as long as standardized ways of providing contact information such as security.txt \cite{draft-foudil-securitytxt-09} are not widely used, it may be possible to ask notification recipients for voluntary contributions to help to cover them. Several recipients expressed gratitude for our notifications, and one museum sent us an (unsolicited) gift of two tickets for their exhibition, indicating that at least some recipients may have been willing to financially support such notifications, if asked. Maass \emph{et al.} reported similar reactions \cite{Maass2021}. An investigation into the willingness to pay for unsolicited notifications (and the impact on remediation of making such a request) is a promising avenue for future work, although any such study would be well-advised to consider the legal implications, in particular the relevant competition and advertising laws in their jurisdiction.
Finally, since our study was performed with German websites only, more research is needed to determine if international notifications benefit from alternative contact channels.
\section{Conclusion}
\label{sec:conclusion}
In this paper, we reported on a randomized controlled notification experiment with 1359 German website operators affected by a set of unintentional information leak vulnerabilities. We compared the effectiveness of emails to that of an alternative notification medium---letters---and the inclusion of more detailed scenarios illustrating the danger of the vulnerability.
We utilized manually-collected address information, finding greatly reduced bounce rates compared to previous studies, with less than 2\,\% of messages being returned as undeliverable.
Overall, \allnotifiedmedian{}\,\% of notified recipients remediated within one month, compared to \controlmedian{}\,\% for the control group.
Letters achieved a substantial increase in remediation rate compared to emails, with differences of up to 25 percentage points.
However, including a more detailed description of the risk posed by the vulnerability not only failed to improve remediation for the email groups, but actually \emph{reduced} remediation rates for letters in some cases.
This counterintuitive result highlights that more work is needed to understand how recipients perceive unsolicited notifications.


\section*{Code and Data}
To facilitate reproduction of our work, we release the code and (anonymized) data necessary to reproduce the figures and tables in this paper, as well as the code of the collection system. Find the data at \url{https://zenodo.org/record/4817464}.
\begin{acks}
This work has been co-funded by the DFG as part of project C.1 within the RTG 2050 “Privacy and Trust for Mobile Users" and by the German Federal Ministry of Education and Research and the Hessen State Ministry for Higher Education, Research and the Arts within their joint support of the National Research Center for Applied Cybersecurity ATHENE.
\end{acks}

\bibliographystyle{ACM-Reference-Format}
\bibliography{bibliography}

\appendix
\section{Example Notification}
\label{appx:notification}
\begin{figure*}
 \includegraphics[width=0.9\linewidth]{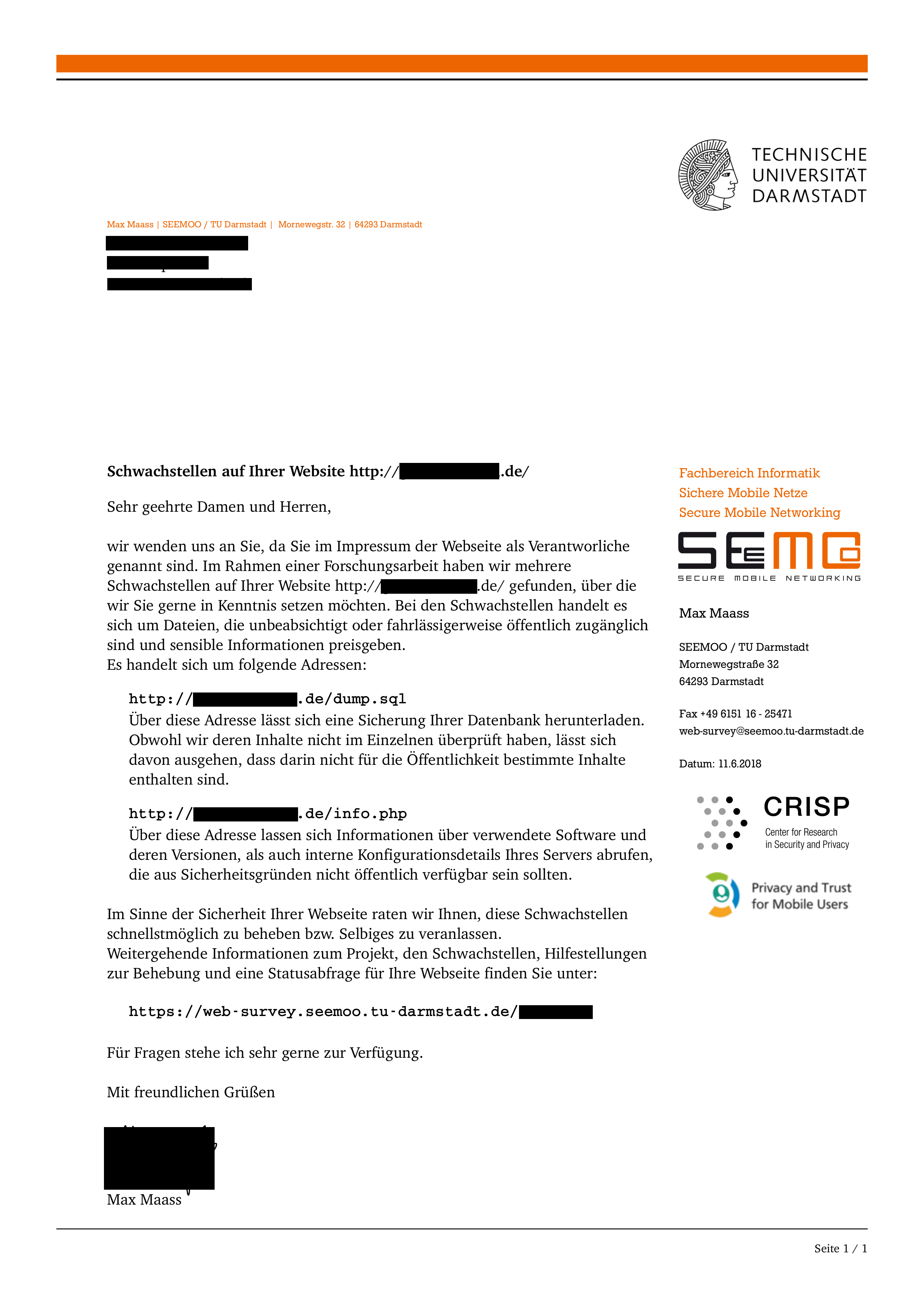}
 \caption{German example notification letter (translation provided below)}
 \label{fig:appx:notification}
 \Description{An example letter showing the letterhead and German content of the letters sent as part of the study. The message is printed on the official letterhead of the university, including the logos of university and projects. The content is reproduced in a translated form below.}
\end{figure*}

\begin{figure*}
	\centering
	\subfloat[Full Groups]{{\includegraphics[width=.45\linewidth]{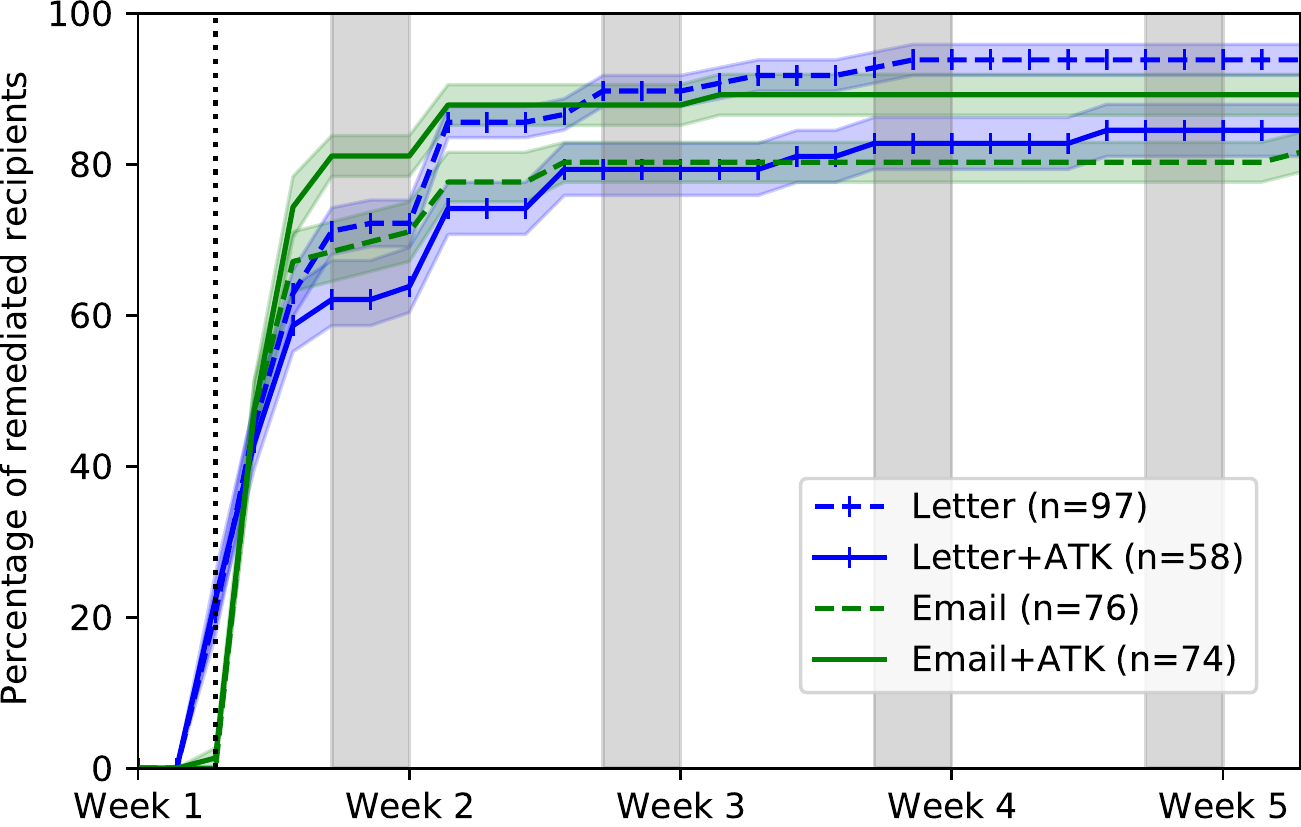}}}%
	\qquad
	\subfloat[Vulnerabilities (Cryptographic Keys omitted for readability)]{{\includegraphics[width=.45\linewidth]{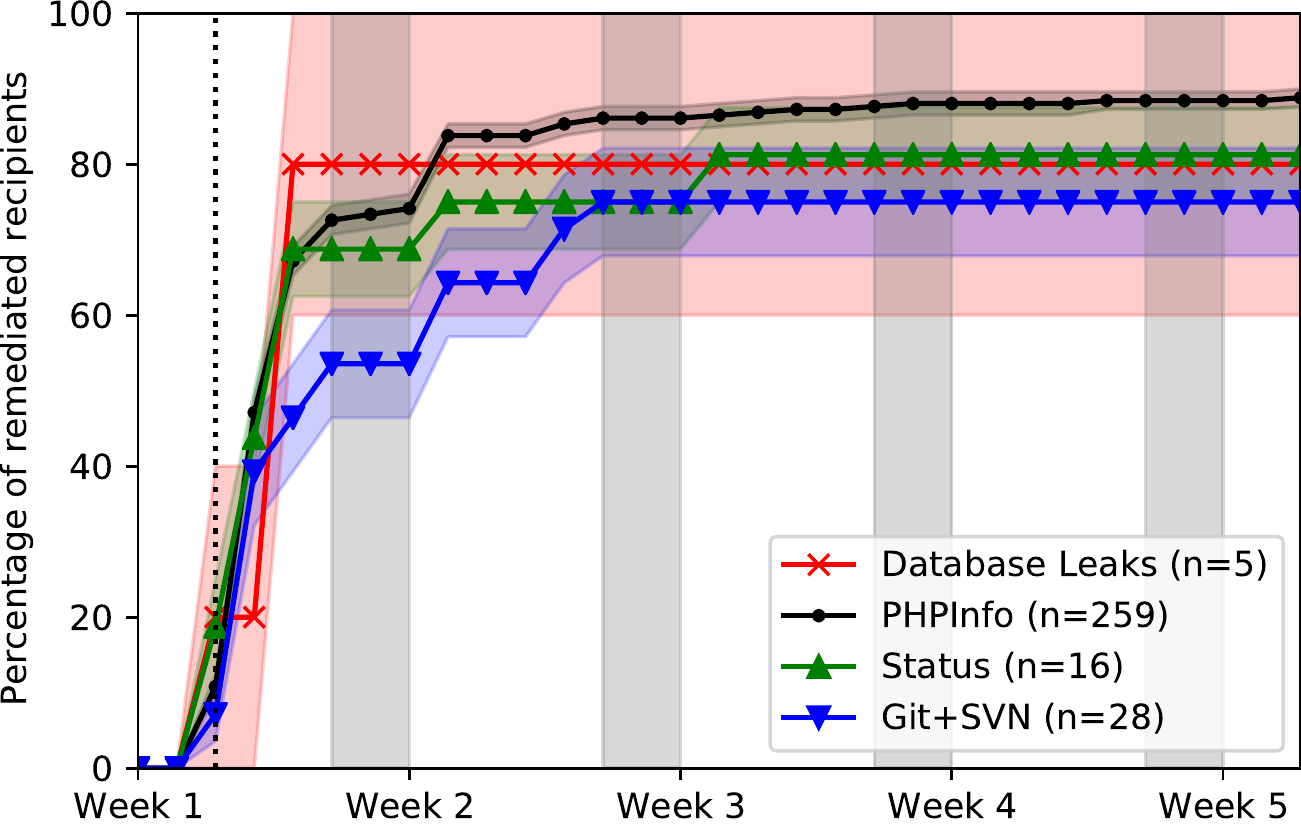}}}\\
	\subfloat[Medium]{{\includegraphics[width=.45\linewidth]{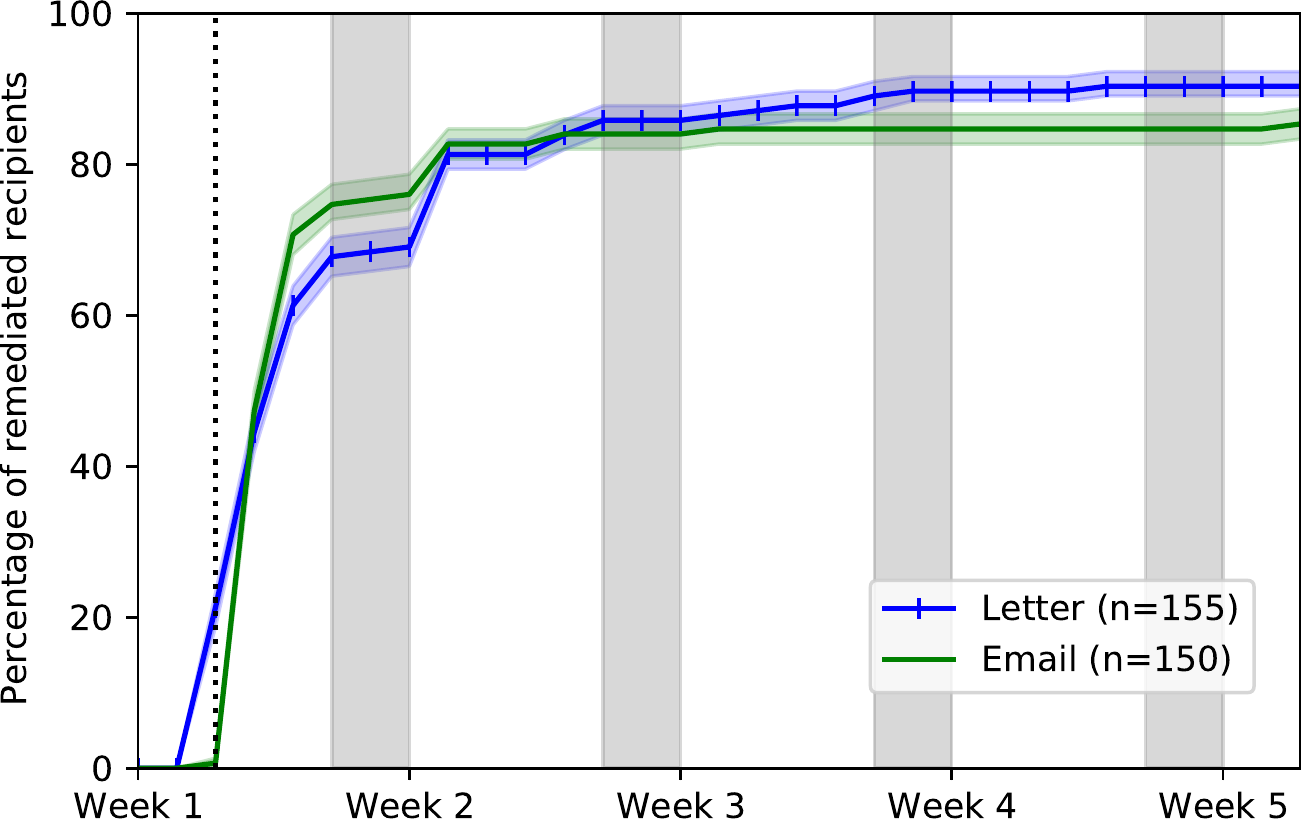}}}%
	\qquad
	\subfloat[Scenario]{{\includegraphics[width=.45\linewidth]{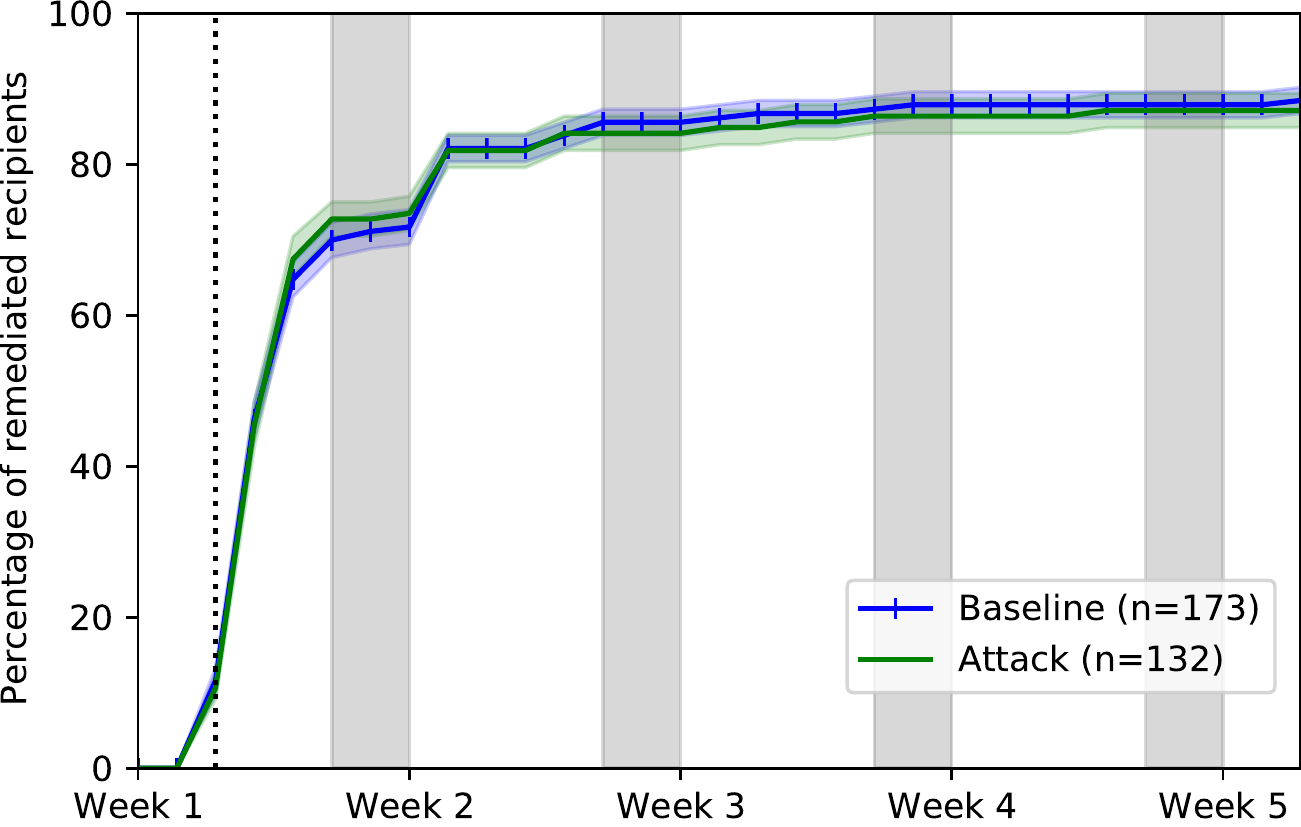}}}
	\caption{Median, 1st and 3rd quartiles for the remediation rates in different experimental groups, vulnerabilities, mediums, and presence of attack scenarios, for reached recipients only.}
	\label{fig:groupcomparison_reached}%
	\Description{A set of four plots. The top left plot shows all combinations of medium and message type. The lines are ordered, from best to worst, as \letter{}, \mailplus{}, \letterplus{}, \email{}. The top right plot shows the different vulnerability types. Ordered from highest to lowest: PHPinfo, Status, Database Leaks, and Git+SVN. The bottom left shows the two different mediums. Letter initially starts out below email, but the lines cross after about 1.5 weeks. In the end, letters slightly outperform emails. The bottom right plot shows the comparison between presence and absence of the attack scenario. The lines are indistinguishable, no differences are visible.}
\end{figure*}

\autoref{fig:appx:notification} shows an anonymized example letter in the original German version. In the following, we will provide translated versions of the explanations of the different vulnerability types that were used in the letter. All messages started with the following text:

\begin{quotation}
  To Whom it may concern,

we are contacting you because you are listed as the responsible person for the website in its imprint. Within the context of a research project we found multiple vulnerabilities on your website
http://www.example.com about which we would like to kindly advise you. The vulnerabilities stem from files that are publicly accessible, either unintentionally or by carelessness, and reveal sensible information.
The following addresses are affected:
\end{quotation}

It was followed by a list of URLs and the relevant explanations, listed below. The message then closed by stating:

\begin{quotation}
In the interest of your websites security we advise you to remedy
those vulnerabilities as soon as possible.
Further information on our project, your vulnerabilities,
assistance in remediation and a status check for your website is available at: [URL with token]

I will be happy to assist you with any further questions.

With kind regards,

Researcher Name
\end{quotation}

The message signature contained the name, fax number, email and postal address of the sender. We will now provide the individual text blocks that describe the different vulnerabilities.

\subsection{SSH Key}
\paragraph{Baseline} At this address anyone can download an access key which can presumably be used to log in to your website and get full access. Please consult an expert for the next steps, because we can not determine the full impact of this problem.

\paragraph{Attack} This vulnerability was only observed once and thus does not contain a version with attack scenario.

\subsection{TLS Key}
\paragraph{Baseline} At this address anyone can access the private key to your websites transport encryption. Because of that we have to assume that the encryption can actually be abrogated. Please deactivate the access to the key, renew the key as well as your encryption certificate and revoke the old key.

\paragraph{Attack} This vulnerability was only observed once and thus does not contain a version with attack scenario.

\subsection{Database Backup}
\paragraph{Baseline} At this address, anyone can download a backup of your database. Although we did not review its content in detail, it is very likely that its contents are not intended for public consumption.

\paragraph{Attack} An attacker can likely extract the list of users from that backup and can possibly extract the passwords. This may grant them full access to the website and allow them to manipulate the content, which can lead to defamation.

\subsection{VCS}
\paragraph{Baseline} The availability of this file indicates that the source code of your website is publicly accessible. While we did not verify those contents in detail, we assume that it contains content that is not meant to be publicly accessible, such as login and contact data or internal configuration files.

\paragraph{Attack} Depending on whether an attacker discovers information like login or contact data, she can in some circumstances acquire full access to your website or impersonate you with the help of the contact data to gain access to further information by fraud.

\subsection{Server-Status}
\paragraph{Baseline} At this address anyone can see which pages on the website are currently accessed from which IP~address using which parameters. This means that in doing so you illicitly disclose the identity and activities of your visitors.

\paragraph{Attack} Using this information an attacker can trace who visits your website and learn about visit duration and links clicked. In the worst case she thereby learns the so called ``Session ID'', which enables her to seize the role of a visitor and impersonate them.

\begin{table}
	\caption{Requested Paths for the different vulnerabilities}
	\begin{tabular}{ll}
		\toprule
		\textsc{Vulnerability} & \textsc{Paths} \\
		\midrule
		Keyfile & \texttt{id\_rsa, .ssh/id\_rsa,}\\
				& \texttt{privatekey.key, private.key,}\\
				& \texttt{myserver.key, key.pem,}\\
				& \texttt{privkey.pem, [domain].key,}\\
				& \texttt{[domain\_full].key, [subdomain].key,}\\
				& \texttt{[domain].pem, [full\_domain].pem,}\\
				& \texttt{[subdomain].pem, cert.pem,}\\
				& \texttt{certificate.pem, domain.key}\\
		Database & \texttt{dump.db, dump.sql, sqldump.sql,}\\
				& \texttt{sqldump.db, db.sqlite, data.sqlite,}\\
				& \texttt{sqlite.db, [domain].sql,}\\
				& \texttt{[domain\_full].sql, [subdomain].sql,}\\
				& \texttt{[domain].db, [domain\_full].db,}\\
				& \texttt{[subdomain].db}\\
		Core dump & \texttt{core} \\
		\gls{VCS} & \texttt{.git/HEAD, .svn/wc.db} \\
		Status & \texttt{server-status/, server-info/} \\
		\phpinfo & \texttt{phpinfo.php, test.php, info.php} \\
		\bottomrule
	\end{tabular}
	\label{tab:vulnpaths}
\end{table}

\subsection{Server-Info}
\paragraph{Baseline} At this address anyone can retrieve information about software modules in use as well as their versions and internal configuration details of your server which should not be public for security reasons.

\paragraph{Attack} With the help of such version information an attacker can very easily determine whether outdated software with known vulnerabilities is in use. If this is the case, she can exploit those easily and in the worst case gain access to the server.

\subsection{PHPInfo}
\paragraph{Baseline} At this address anyone can access information about the software in use as well as internal configuration details, which should not be publicly accessible for security reasons.

\paragraph{Attack} With the help of such version information an attacker can easily determine whether outdated software with known vulnerabilities is in use. If this is the case, she can exploit them with ease and in the worst case gain access to the server.


\end{document}